\documentclass[aps,prb,reprint,superscriptaddress]{revtex4-2}
\usepackage{hyperref}
\usepackage{graphicx}
\usepackage{color}
\usepackage{amsfonts}
\usepackage{amsmath}
\usepackage[utf8]{inputenc}
\usepackage[T1]{fontenc}
\usepackage{mathtools}
\usepackage{bbm}
\usepackage[dvipsnames]{xcolor}
\hypersetup{colorlinks=true, urlcolor=blue, citecolor = blue}

\begin{document}

\title{Nonlocal transport signatures of topological superconductivity\\ in a phase-biased planar Josephson junction}

\author{D. Kuiri}
\email{kuiri@agh.edu.pl}
\affiliation{AGH University of Krakow, Academic Centre for Materials and Nanotechnology, al. A. Mickiewicza 30, 30-059 Krakow, Poland}

\author{M. P. Nowak}        
\email{mpnowak@agh.edu.pl}
\affiliation{AGH University of Krakow, Academic Centre for Materials and Nanotechnology, al. A. Mickiewicza 30, 30-059 Krakow, Poland}

\date{\today}

\begin{abstract}
Hybrid Josephson junctions realized on a two-dimensional electron gas are considered promising candidates for developing topological elements that are easily controllable and scalable. Here, we theoretically study the possibility of the detection of topological superconductivity via the nonlocal spectroscopy technique. We show that the nonlocal conductance is related to the system's band structure, allowing probe of the gap closing and reopening related to the topological transition. We demonstrate that the topological transition induces a change in the sign of the nonlocal conductance at zero energy due to the change in the quasiparticle character of the dispersion at zero momentum. Importantly, we find that the tunability of the superconducting phase difference via flux in hybrid Josephson junctions systems is strongly influenced by the strength of the Zeeman interaction, which leads to considerable modifications in the complete phase diagram that can be measured under realistic experimental conditions.
\end{abstract}

\maketitle

\section{Introduction}
Planar superconductor-normal-superconductor (SNS) Josephson junctions have been proposed as a promising platform for engineering and exploiting Majorana bound states due to the tunability of the topological transition by the superconducting phase difference \cite{PhysRevX.7.021032} and the scalability of 2D heterostructure systems \cite{PhysRevLett.118.107701}. For the realization of topological SNS devices, typically two separate superconducting electrodes proximitize the two-dimensional electron gas (2DEG), creating a SNS junction where the good quality of the normal-superconducting interfaces results in an induced gap close to that of the parent superconductor \cite{Kjaergaard2016}. Upon application of an in-plane magnetic field, the Zeeman interaction leads to the splitting of Andreev bound states (ABS) in phase, resulting in the opening of the topological regime whenever the fermion parity is odd \cite{PhysRevX.7.021032}. The topological regime is already obtained for the vanishingly small Zeeman interaction energies at the phase difference $\pi$. This becomes an important factor in achieving the topological superconductivity in SNS junctions realized on normal-superconductor hybrids as a strong Zeeman interaction can lead to the appearance of abundance of trivial in-gap states \cite{PhysRevLett.119.176805,Moehle2021} that decrease the induced gap and can obscure Majorana zero-energy modes.

Originally, in normal-superconductor nanostructures, such as proximitized nanowires \cite{Zhang2017}, Majorana bound states were sought by tunneling spectroscopy, where the presence of a zero-bias peak was assigned to the appearance of topological zero-energy states \cite{doi:10.1126/science.1222360, Das2012, Deng2012}. However, zero-bias peaks can also result from disorder-induced trivial ABS \cite{Das_Sarma_disorder, PhysRevResearch.2.013377, PhysRevB.103.214502} or be due to the specifics of the tunneling barriers used \cite{10.21468/SciPostPhys.7.5.061}.

Tunneling spectroscopy was considered \cite{PhysRevB.102.245403} (among other techniques such as scanning tunneling microscopy \cite{PhysRevB.102.085414} or its spin-polarized variant \cite{PhysRevB.102.085411}) as a way to probe bound states in planar SNS junctions. It allowed revealing the edge-dependent evolution of ABS in the perpendicular field \cite{PhysRevLett.124.226801, Moehle2022, PhysRevLett.130.116203}. In fact the zero-bias peaks were observed in planar SNS junctions \cite{Fornieri2019, PhysRevLett.130.096202} but, as in the case of nanowire systems, single-edge conductance cannot be considered as a conclusive determinant of the topological character of a zero-energy state. Relying on the observation of a single feature expected for topological systems can lead to false positive results \cite{frolov2023smoking, PhysRevResearch.2.013377}. Instead, specialized protocols \cite{pikulin2021protocol,PhysRevB.106.075306} have recently been proposed that require the observation of several signatures of the topological transition, preferably in a wide range of experimentally controllable parameters.


In this context, a promising method is a nonlocal measurement \cite{PhysRevB.103.014513,PhysRevB.97.045421}, which has recently been the subject of intense research effort. Local and nonlocal spectroscopy was recently performed on planar SNS junctions in the tunneling regime \cite{PhysRevLett.130.096202}, but lacked a clear signature of the topological transition. In this work, we focus on the signatures of the topological transition in phase-biased SNS junctions in nonlocal measurements in the context of recently realized two-dimensional Josephson junction heterostructures \cite{PhysRevB.97.045421,PhysRevLett.124.226801,Moehle2022}. We theoretically study the transport features of the junction both in the spectroscopy limit, i.e. tunneling measurements that are sensitive to the density of states in the junction and in an open regime, i.e., without tunneling barriers, where the transport features correspond rather to a band structure of the junction. The latter can elucidate the closing and reopening of the gap associated with the topological transition \cite{PhysRevB.97.045421}.

We find that the nonlocal conductance sign represents the electron- or holelike character of the bands in the junction. The closing and reopening of the gap at $k = 0$ is associated with the meeting of the electron and hole bands at zero energy and, correspondingly, with the change in the sign of the nonlocal conductance. Furthermore, we discuss a serious caveat in the realization of the topological phase in SNS junctions. Namely, we show that in a realistic situation, where phase biasing is done by running a flux through a superconducting loop embedding the SNS junction, the phase slips result in skipping a large region of phase space close to $\pi$, prohibiting the creation and probing of Majorana bound states at a small field. We discuss the factors that allow to limit this obstacle.

The paper is structured as follows. In Sec. II we introduce the numerical model. In Sec. III A we discuss nonlocal spectroscopy results in relation to the effective charge polarization of the bands. In Sec. III B we show how experimentally performed phase biasing limits the magnetic fields in which the topological phase can be observed. We discuss our results in Sec. IV and summarize them in Sec. V. 

\section{Model}
\begin{figure}[ht!]
    \centering
    \includegraphics[width = 0.45\textwidth]{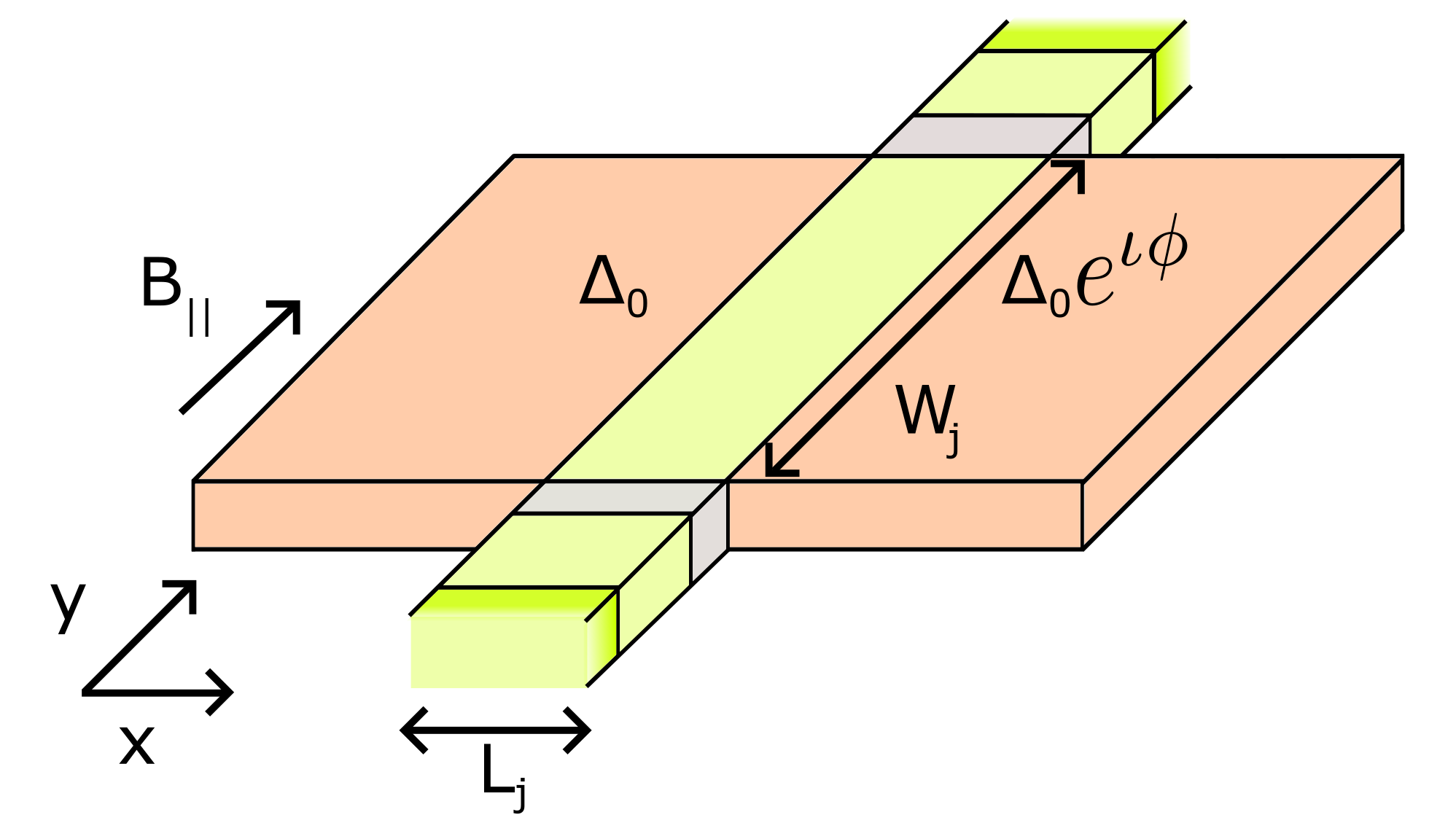}
    \caption{Schematic diagram of the considered system. A semiconductor strip (yellow-green) is sandwiched between two superconducting electrodes (orange), whose pairing potential has phase difference $\phi$. The gray regions denote the potential barriers, placed just above and below the SC region. The green segments in the semiconductor denote the top (1) and bottom (2) normal leads.}
    \label{fig:system}
\end{figure}

We consider a planar SNS junction constituted by a semiconducting strip of length $L_j$ connected to superconducting electrodes of width $W_j$. The scheme of the considered structure is depicted in Fig. \ref{fig:system}.

The Hamiltonian of the system written in the basis $\Psi = (\psi_{e\uparrow}, \psi_{h\downarrow}, \psi_{e\downarrow}, -\psi_{h\uparrow})^T$ (where $e$ and $h$ correspond to electron and hole components with spin up $\uparrow$ or down $\downarrow$ respectively) is
\begin{equation}\label{eq:1}
\begin{split}
       H &= \left( \frac{\hbar^2 {k_x}^2}{2m^*} + \frac{\hbar^2 {k_y}^2}{2m^*} - \mu \right)\sigma_0 \otimes \tau_z + \frac{1}{2}g(x)\mu_{B}B\sigma_y \otimes \tau_0 \\
       &\quad + \alpha(x)(\sigma_x k_y - \sigma_y k_x)\otimes\tau_z + \Delta(x) \tau_{+} +  \Delta^{*}(x) \tau_{-}.
\end{split}
\end{equation}
where $k_{x(y)} = -\iota \partial/\partial x(y)$, $\sigma_i$ and $\tau_i$ with ($i = x, y, z$) are the Pauli matrices that act on the spin and electron-hole degree of freedom, respectively, with $\tau_{\pm} = (\sigma_0\otimes\sigma_x \pm \iota \sigma_0\otimes\sigma_y)/2$ where $\sigma_0$ is ($2 \times 2$) identity matrix.

We consider the nonzero pairing potential in superconducting contacts, which is modeled by the spatial dependence of the gap parameter $\Delta(x)$, 
\[ \Delta(x) = \begin{cases*}
                    \Delta_{0} & if  $x < -L_j/2$  \\
                     \phantom{}0 & if $-L_j/2\leq x \leq L_j/2$\\
                     \phantom{}\Delta_{0}e^{\iota\phi} & if $x > L_j/2$,
                 \end{cases*} \]%
with $\phi$ the superconducting phase difference. Accordingly, we neglect the Zeeman splitting and spin-orbit effects in the superconductor setting $g(x)=\alpha(x)=0$ in them. 
The in-plane magnetic field is applied along the $y$ direction. For concreteness, we adopt the material parameters corresponding to the InSb semiconductor and the Al superconductor, that is, $m^* = 0.014m_e$, $\mu = 5$~meV, $\Delta_0 = 0.2$~meV, $\alpha = 50$~meVnm. We also assume typical dimensions for this type of structure, that is, $L_j = 80$ nm, $W_j = 2000$ nm \cite{Moehle2022, PhysRevLett.130.096202,PhysRevB.107.245304}. 

For numerical simulations, we discretize the Hamiltonian on a square lattice with the lattice constant $a = 10$~nm. Since we use a uniform chemical potential in our calculations, for a proper description of Andreev scattering at the NS interface \cite{PhysRevB.59.10176}, we introduce an anisotropic mass in the superconducting leads with the effective mass in the direction parallel to the interface $m^*_\parallel = 10 m^*$ \cite{short_junctions}.  The code used for the presented calculations is available online \cite{zenodo_repository}.

In this study, we consider three variants of the SNS system. The first is the {\it open} system as shown in Fig. \ref{fig:system} used to study the transport properties. Here, the normal regions extend beyond the width of the superconducting contacts by length 100 nm, which includes 10 nm tunneling barriers of height 50 meV. They are connected to semi-infinite leads that allow the transport of in-gap electrons/holes into/from the junction. The normal region is connected to semi-infinite superconducting leads. Such geometries have recently been experimentally realized in InSbAs \cite{Moehle2022} or InAs \cite{PhysRevLett.130.096202} 2DEGs proximitized by Al.

Experimentally, the nonlocal conductance is measured by grounding the superconductor and registering the current change in one of the normal leads upon application of the voltage bias in the other. Numerically we calculate the nonlocal conductance considering the scattering properties of the quasiparticles injected and scattered back to normal leads using the scattering matrix approach implemented in the Kwant package~\cite{kwant} with the formula
\begin{equation}
\label{eq:3}
    G_{ij}(E) = \frac{\partial{I_i}}{\partial{V_j}} = \frac{e^2}{h} (T^{ee}_{ij} - T^{he}_{ij} - \delta_{ij} N^e_i).
\end{equation}
$I_i$ is the current entering terminal $i$ from the scattering region, $V_j$ is the voltage at terminal $j$ and $N^e_i$ is the number of electron modes at energy $E$ in terminal $i$. $T^{ee}_{ij}$ and $T^{he}_{ij}$ are electron-to-electron and electron-to-hole transmission amplitudes (with $j$ being the source and $i$ the drain) calculated at energy $E$ that represents the applied bias voltage $V_j$ at zero temperature \cite{PhysRevB.97.045421}.

To investigate the properties of the bound states that form in the junction, we consider a finite {\it isolated} system by disconnecting the protruding normal segments and leads \cite{SC_leads_note}. Finally, to study the properties of the band structure, we introduce the {\it translation-invariant} system constructed by removing the normal leads and making the junction invariant in the $y$ direction, where $k_y$ is a good quantum number.

\section{Results}
\subsection{nonlocal conductance as a measure of topological transition}
\begin{figure}[ht!]
    \centering
    \includegraphics[width = 0.4\textwidth]{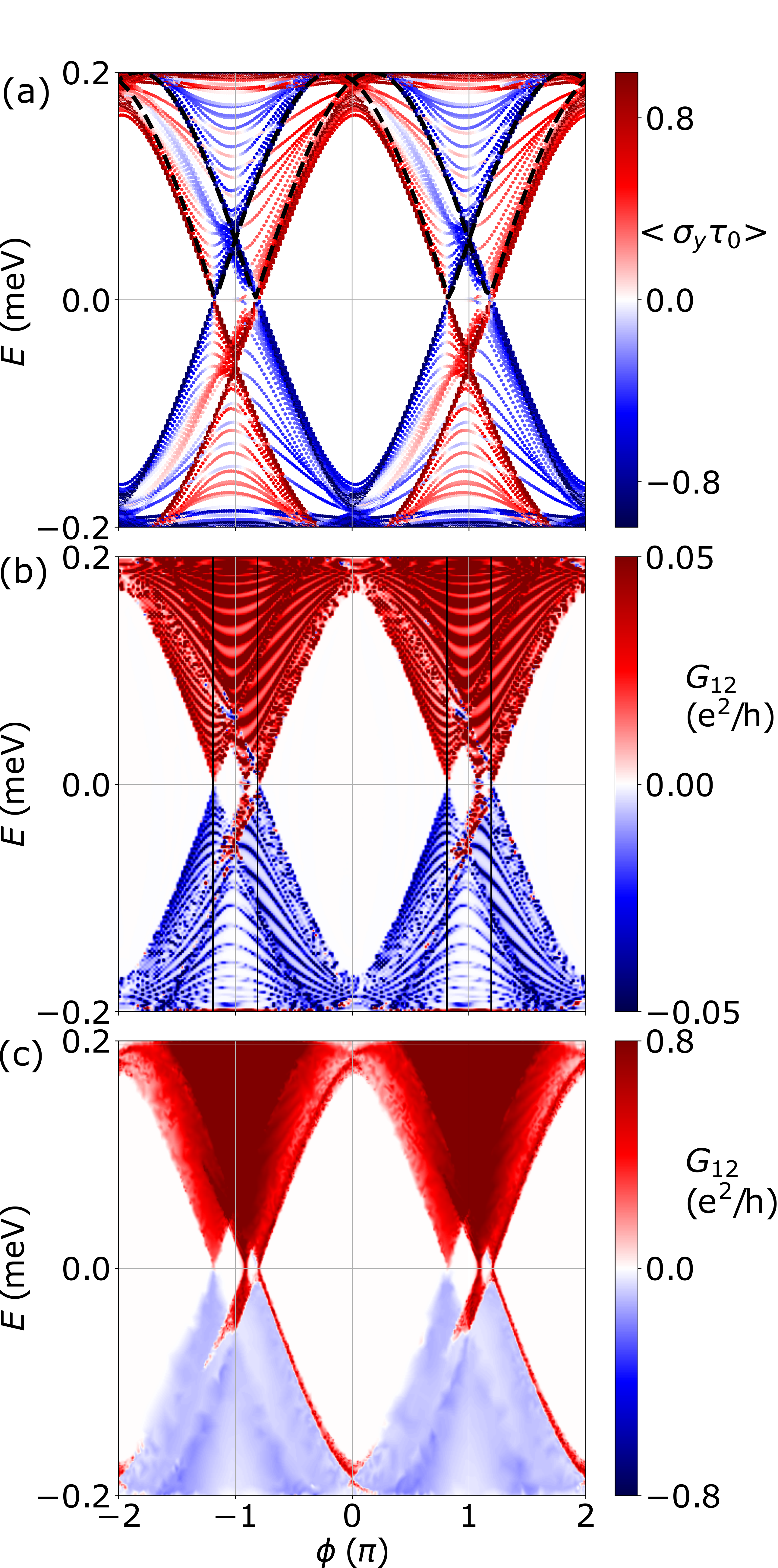}
    \caption{(a) ABS energy spectrum of the isolated SNS junction. The analytical ABS spectrum is shown with black dashed lines. The colors denote the average spin polarization of ABS along the $y$ direction. Nonlocal conductance with (b) and without (c) tunneling barriers versus the superconducting phase difference. The results are obtained for the in-plane field $B = 0.5$ T.}
    \label{fig:abs}
\end{figure}

The $2\pi$-periodic spectrum of isolated junction in a non-zero in-plane field is shown in Fig. \ref{fig:abs}(a). The evolution of the ABS of a single-mode spinful junction in the presence of the magnetic field is captured by the formula
\begin{equation}
E_\sigma(\phi) = \Delta \sqrt{1- \tau \sin^2\left(\frac{\phi + \varphi_\sigma}{2}\right)},
\label{ABS_energy}
\end{equation}
where $\varphi_\sigma = 2\sigma E_{z}L_j/\hbar v_{F}$, $\tau$ is the junction transmission coefficient, $E_z = g\mu_{B}B/2$ is the Zeeman energy, $\sigma = \pm 1$ corresponds to positive and negative spin components and $v_F = \sqrt{2\mu/m^*}$ is the Fermi velocity. Overlaying the numerically calculated spectrum with the analytical one in Fig. \ref{fig:abs}(a) we see that the cones made up of ABS are split in phase by the Zeeman interaction. Inspecting the mean spin polarization of ABS calculated as the expectation value of the operator $\sigma_y\tau_0$, we observe that the edge of each cone is made of ABS with positive and negative spin polarization along the $y$ direction. Upon increasing the magnetic field, for the negative $g$ factor considered here, the positively (negatively) spin-polarized states move down (up) in energy. This in turn results in an increase in the distance between the positive-energy cones. The bottom tip of each cone sets out a phase point when the fermion parity changes and the system undergoes a phase transition---with the topological phase being present in each $\phi = [0, 2\pi]$ (mod $2\pi$) segment only between the cones. Since the plot shows spin polarization, the spinless Majorana bound states are not visible.

Calculating the nonlocal tunneling spectroscopy we obtain the map shown in Fig. \ref{fig:abs}(b) where the gap closing and reopening upon the increase of the superconducting phase is visible. An analogous result is obtained when the tunneling barriers are removed [Fig. \ref{fig:abs}(c)]. Most importantly, we observe that in both plots the topological transition manifests itself as the sign change of the nonlocal signal at zero energy [see the vertical black lines in Fig. \ref{fig:abs}(b)] leading to the rectification of the current, similar to the case of an NS junction \cite{PhysRevB.97.045421}.

In Figs. \ref{fig:abs}(b) and \ref{fig:abs}(c) there is also a middle cone visible in which the nonlocal conductance does not change sign at zero energy, and which does not mark the topological transition, as we will show in the following.

\subsubsection{Charge polarization of the bands}
As we will show, the sign of nonlocal conductance outlines the leading transport phenomenon in the junction. According to the formula, Eq. (\ref{eq:3}) a positive conductance signal is obtained when the dominant transport process involves electron transport through the proximitized region, while a negative signal is obtained when the electron is converted into a hole in a crossed Andreev reflection process.

\begin{figure}[ht!]
    \centering
    \includegraphics[width = 0.35\textwidth]{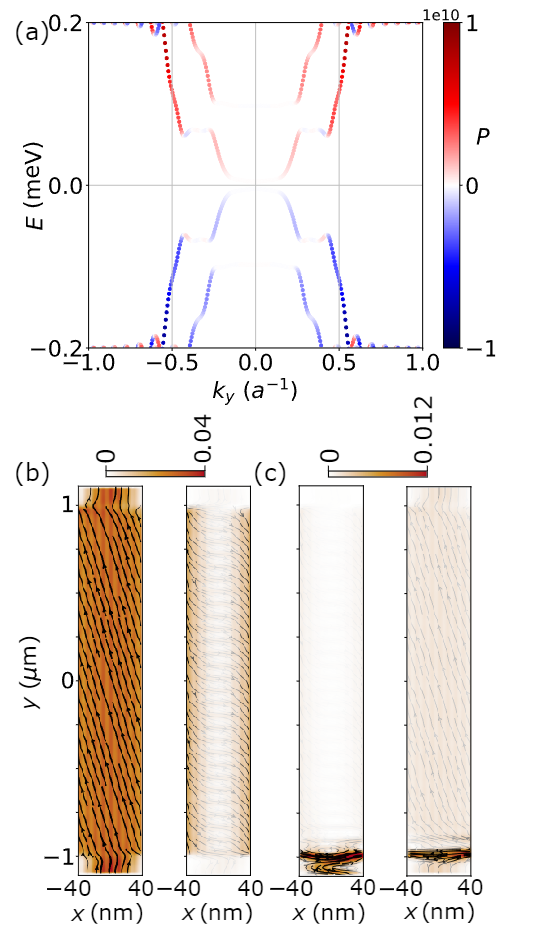}
    \caption{(a) Dispersion relation for $B=0.5$ T and $\phi = 0.84\pi$. The colors denote the average charge polarization of the bands. Electron (left) and hole (right) components of probability currents obtained for $E = 0.169$ meV (b) and $E = -0.169$ meV (c).}
    \label{fig:dispersion}
\end{figure}

To elucidate the change of the nonlocal conductance, we consider an invariant system. In Fig. \ref{fig:dispersion}(a) we plot the dispersion relation obtained for the phase difference set in the vicinity of the left cone, that is, $\phi = 0.84\pi$. We see the gap closing at $k_y = 0$ as the bands cross zero energy, causing the fermion parity change, which in turn leads to the topological transition. 

We introduce the quasiparticle polarization of the bands factor ($P$) which is calculated as $P = vk_{y}$, where $v = \frac{1}{\hbar}\frac{\partial E}{\partial k_y}$ and color the bands in the dispersion relation in Fig. \ref{fig:dispersion}(a) with it. We observe that the bands at positive energy mostly have an electron-like character, i.e. the sign of the Fermi velocity matches the sign of the wave vector $k_y$. The situation for negative energy is the opposite, and the bands there are mostly of a hole-like character. 

Positive band polarization allows electrons with positive energies to flow between the top and bottom contacts with little Andreev reflection [see Fig. \ref{fig:dispersion}(b)]. We observe that the electron can freely propagate from the bottom to the top contact [left map in Fig. \ref{fig:dispersion}(b)] with little Andreev reflection [right map in Fig. \ref{fig:dispersion}(b)]. On the other hand, the mostly hole character of the negative energy bands results in a blockade of the electron transport [see the left map in Fig. \ref{fig:dispersion}(c)], and instead crossed Andreev reflection occurs [see the right map in Fig. \ref{fig:dispersion}(c)],  which in turn results in negative nonlocal conductance that is related to the splitting of the Cooper pairs between top and bottom contacts.

It should also be noted that part of the band structure can also have an electron-like character, even at negative energy, which, e.g., can cause the red outlines of the rightmost cone at negative energy as seen in Fig. \ref{fig:abs}(c).

\begin{figure}[ht!]
    \centering
    \includegraphics[width = 0.4\textwidth]{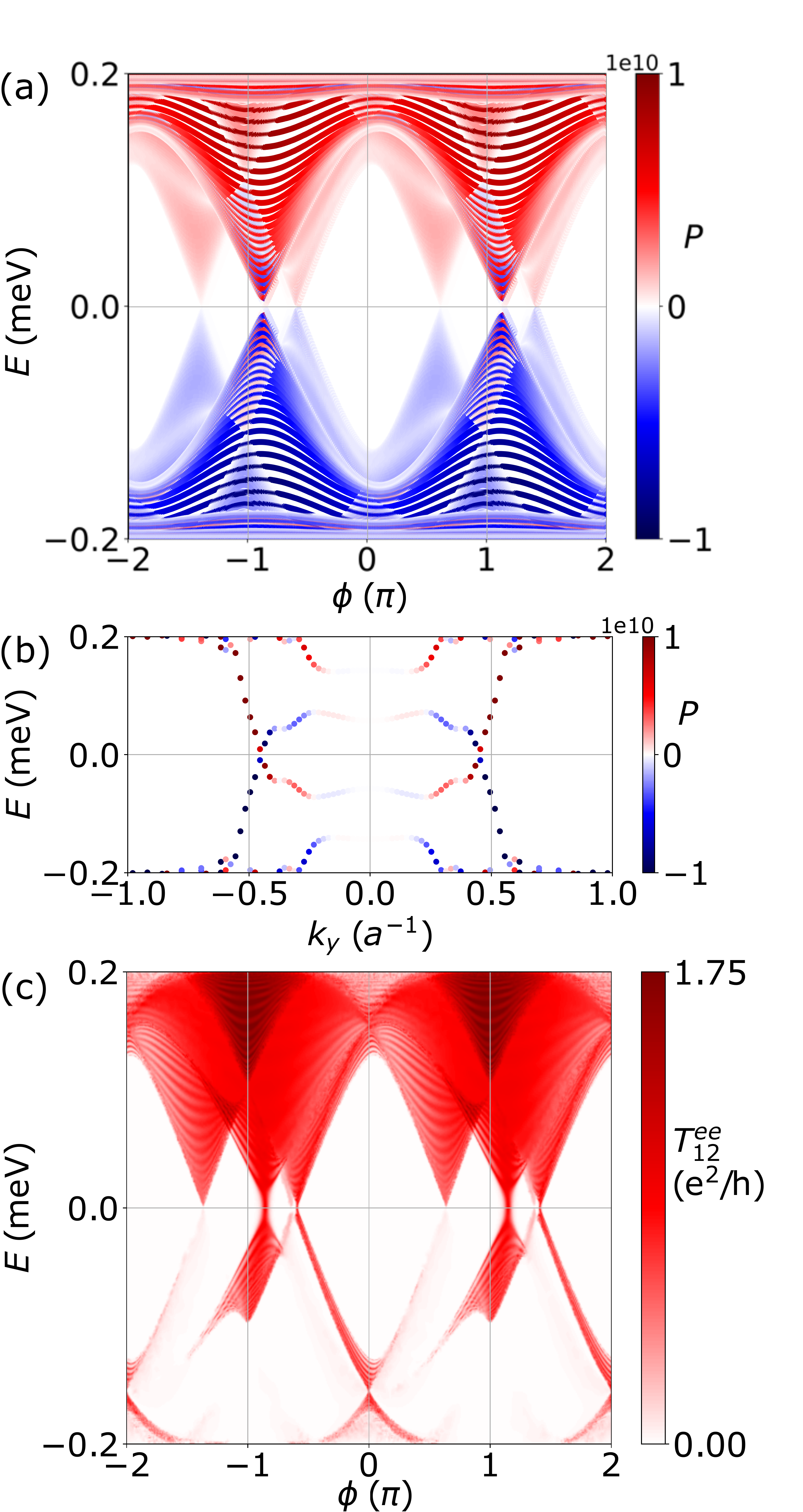}
    \caption{Charge polarization obtained for $ k_y \in [-a^{-1}, a^{-1}]$ versus energy and phase difference. (b) The band structure obtained for $\phi = 1.15\pi$. (c) $T^{ee}_{12}$ nonlocal conductance component. The results are obtained for $B=1$ T.}
    \label{fig:middlecone}
\end{figure}

In Fig. \ref{fig:middlecone}(a) we show the $P$ factor for the invariant system calculated by projecting the $P$ values obtained in the range $k_y \in [-a^{-1}, a^{-1}]$ for each phase difference value for the increased in-plane field $B = 1$ T. We indeed see that in the outermost cones in each $2\pi$ segment of the spectrum the particle polarization of the bands is positive at positive energy and vice versa. The opposite polarizations result in the change of the sign of the nonlocal conductance at zero energy, which marks the topological transition.

In the map of Fig. \ref{fig:middlecone}(a) there is also a clear signature of the appearance of positively charged bands both in positive and negative energy between the two outermost cones in the topological region. If we look at the exemplary dispersion relation, obtained for the phase where the positive- and negative-energy middle cones meet, [Fig. \ref{fig:middlecone}(b)] we observe that the gap closing occurs at non-zero $k_y$. Therefore, this gap closing does not result in a phase transition. The modes in those bands have a high Fermi velocity at zero energy and therefore a small $\phi_\sigma$ Zeeman phase shift that results in a weak dependence on the position of this cone on the strength of the in-plane field. Finally, since those bands always have a considerable electron polarization, the electrons can be transmitted through the system for both positive and negative energies. This is clearly visible in the map of Fig. \ref{fig:middlecone}(c), where we show the electron transmission coefficient. This effect in turn results in lack of sign change of the nonlocal conductance at zero energy as it is for the cones that mark the topological/trivial transition.

\subsection{Phase biasing by a perpendicular magnetic field}
\begin{figure}[ht!]
    \centering
    \includegraphics[width = 0.35\textwidth]{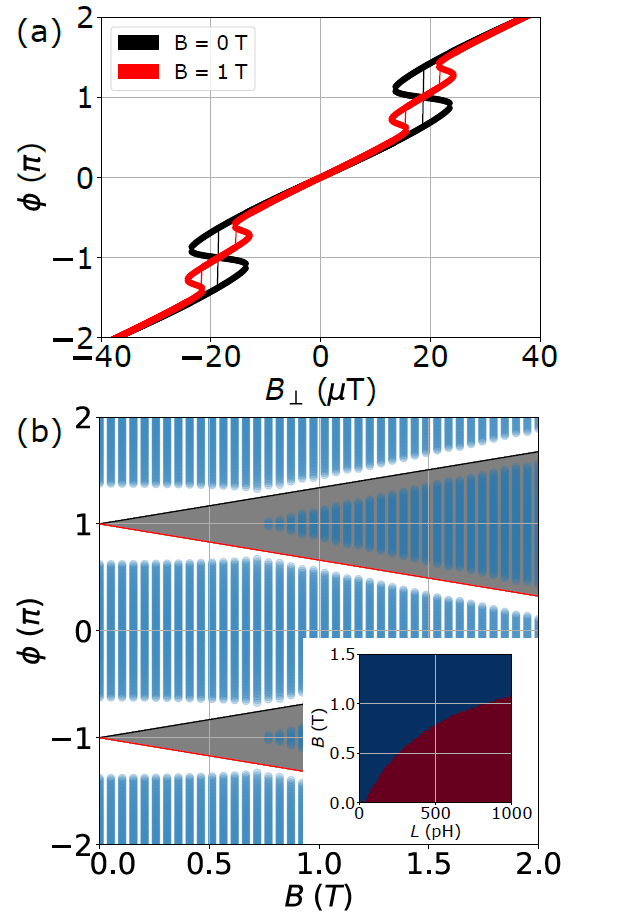}
    \caption{(a) Plot of phase difference versus perpendicular field obtained without (black) and with (red) the in-plane field. The thick curves show the results without energy minimization, and the thin lines correspond to the case of the energy minimization included. (b) The blue dots represent the feasible phase values that can be obtained by applying the perpendicular magnetic field $B_{\perp}$ for a given value of the in-plane field $B$. Gray areas, outlined by black and red lines, show the predicted [from the solution of Eq. \ref{ABS_energy}] phase-magnetic field range in which the topological regime is expected. The inset shows in red the parameter range in which the $\phi = \pi$ phase bias is available upon variation of the superconducting loop inductance.}
    \label{fig:analyticalBtophase}
\end{figure}

Phase biasing of the junction is achieved by placing the junction in a superconducting loop and threading the loop with a perpendicular magnetic field $B_{\perp}$ resulting in the magnetic flux $\Phi = B_{\perp}\pi R^2$, with $R$ being the radius of the loop. Typically, those loops have significant inductance $L$ \cite{Moehle2022,PhysRevB.107.245304} leading to non-linear magnetic field to phase conversion governed by the equation
\begin{equation}
    \phi = \frac{2 \pi}{\Phi_0}\left(\Phi - L \sum_{\sigma = \pm 1} I_\sigma(\phi)\right).
\label{fluxphase}
\end{equation}

Typically the perpendicular field $B_{\perp}$ magnitude is a few orders lower than the magnitude of the in-plane field, therefore one can consider that it does bring negligible effects in terms of Zeeman spin splitting. The Zeeman interaction due to the in-plane field nevertheless leads to the evolution of ABSs through the formula Eq. (\ref{ABS_energy}) and causes the modification of the supercurrent whose phase dependence at zero temperature for a junction embedding $M$ spinful modes can be approximated as
\begin{equation}
I_\sigma(\phi) = \frac{e \Delta^2 \tau M}{4\hbar}\frac{\sin(\phi + \varphi_\sigma)}{E_\sigma(\phi)}.
\label{ianal}
\end{equation}

\begin{figure*}[ht!]
    \centering
    \includegraphics[width = 0.9\textwidth]{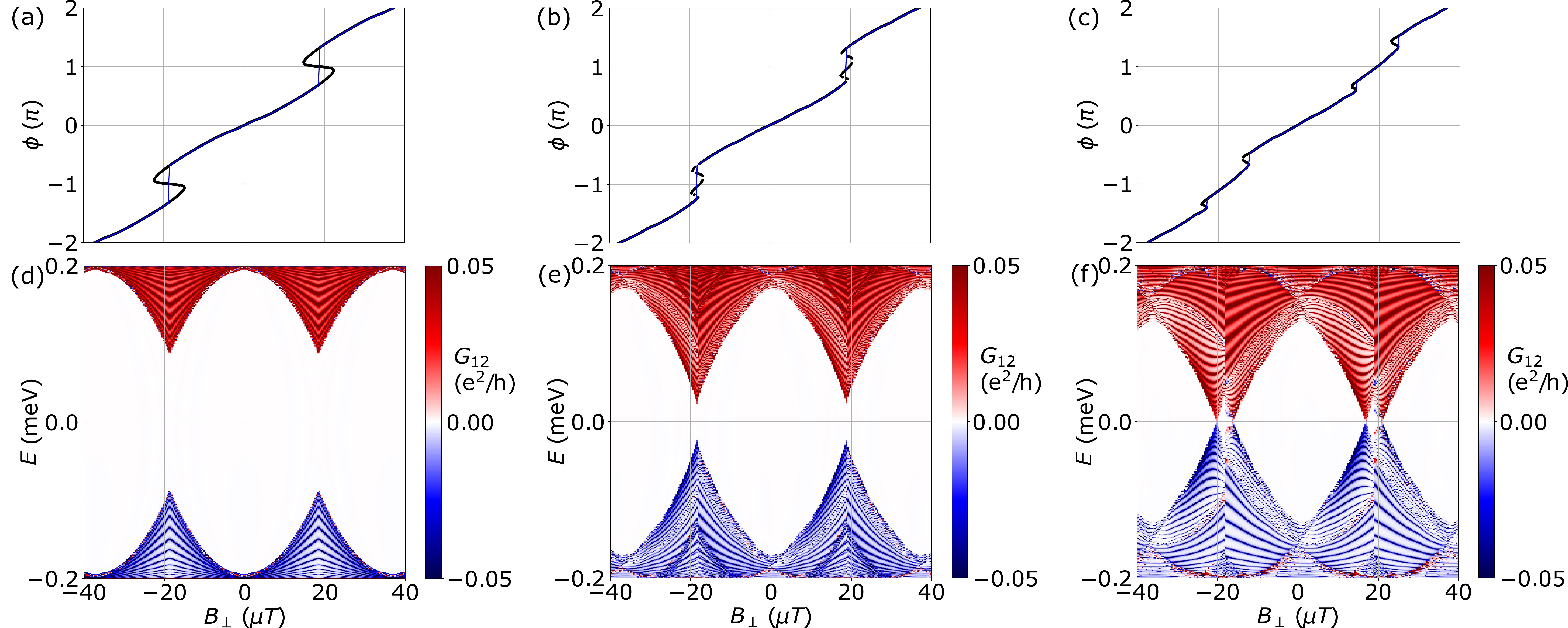}
    \caption{(a),(b),(c) Perpendicular field to phase conversion obtained from numerical ABS spectrum for absent (a) parallel field and (b) 0.5 T
and (c) 1 T. (d),(e),(f) Nonlocal tunneling spectroscopy results as a function of the perpendicular field used for phase biasing }
    \label{fig:spectroscopyinB}
\end{figure*}

The $B_{\perp}$ to phase conversion obtained in the absence of the in-plane magnetic field $B$, where $\varphi_\sigma = 0$, is plotted with a thick curve in Fig. \ref{fig:analyticalBtophase}(a). The obtained dependency is strongly non-linear due to the $LI(\phi)$ term in Eq.~(\ref{fluxphase}). Assuming a quasi-static approximation, for each value of $B_{\perp}$ we minimize $\varepsilon(\phi) = L(\sum_{\sigma = \pm 1}I_\sigma(\phi))^2/2 - M\sum_{\sigma = \pm 1}E_\sigma(\phi)$ to obtain the phase difference that guarantees the ground state of our system. The result is a single-valued conversion curve $B_{\perp}$ to phase presented with a thin black line in Fig. \ref{fig:analyticalBtophase}(a). Here we take the parameters corresponding to the recent experiment \cite{Moehle2022}, i.e., $M = 30$, $L=321$ pH, $\tau$ = 0.99 and $R$ = 4207 nm.

Following the curve from the negative values of $B_\perp$ we observe phase slips close to the values $-\pi$ and $\pi$. As a result, regulating the phase difference by the perpendicular field allows one to obtain phase values only from certain regions \cite{Moehle2022}, which actually omits the most desired values close to $\pi$. The Zeeman interaction leads to a splitting in the ABS structure, as seen in Fig. \ref{fig:abs}(a). As a result, the current jumps are less pronounced and no longer occur at $\pm \pi$---see the red dots in Fig. \ref{fig:analyticalBtophase}(a).

In Fig. \ref{fig:analyticalBtophase}(b) with blue dots we show the possible attainment of phase difference values versus the in-plane magnetic field. We indeed see that only at considerable Zeeman splitting energies it becomes possible to induce the $\pi$ phase difference. In the same plot, we denote the analytical estimate of the phase values that guarantee the topological regime obtained from the analytical ABS spectrum as presented in Fig. \ref{fig:abs}(a). It is clear that, despite the topological gap opening at an already small parallel magnetic field at $\phi = \pi$ it is not possible to set the necessary phase bias to actually induce the topological phase. We observe that only a strong Zeeman interaction unveils the phase-difference region, close to $\pm \pi$---where the topological superconductivity is present. This shows that the Zeeman interaction not only leads to the opening of the topological transition in the junction due to splitting of the ABS but also significantly modifies the flux-phase conversion that is necessary to bias the junction into the topological regime.

Finally, we study the case where the flux-phase conversion is calculated from a numerical spectrum of the junction instead of a simple approximation of Eq. \ref{ianal}. For each value of the in-plane field, we calculate numerically the spectrum of an isolated junction and then obtain the supercurrent $I(\varphi) = -\frac{e}{\hbar} \sum_{E_{n}>0} {\frac{\partial E_n}{\partial \varphi}}$. We then follow the same procedure of flux-to-phase conversion as described above. The results for three values of the in-plane field are shown in Figs. \ref{fig:spectroscopyinB}(a), \ref{fig:spectroscopyinB}(b) and \ref{fig:spectroscopyinB}(c). We see that only for a strong in-plane field the linear flux to phase conversion is restored with the possibility to bias the junction with $\pi$ phase difference.

In Figs. \ref{fig:spectroscopyinB}(d), \ref{fig:spectroscopyinB}(e) and \ref{fig:spectroscopyinB}(f) we show nonlocal spectroscopy results versus the perpendicular field. We see that despite considering a transparent junction the probed ABS do not touch zero energy due to phase slips. Hence the topological region, although present in the spectra plotted against the phase difference in Fig. \ref{fig:abs}(a), is not present when we consider a realistic situation of flux-induced phase biasing. Only at considerable Zeeman interaction strength (here 1T) is the transition between the trivial and topological regime visible as signified by the gap closing and reopening associated with the change of the sign of nonlocal conductance at zero energy [see Fig. \ref{fig:spectroscopyinB}(f)].

\section{Discussion}
We showed that despite the fact that the phase bias can lower the critical Zeeman energy required for the topological transition \cite{PhysRevX.7.021032}, the phase biasing and specifically tuning the junction to $\phi = \pi$ configuration turns out to be difficult to perform in practice, and is in fact dependent on the microscopical parameters of the SNS junction (such as the supercurrent) and the device geometry itself (e.g. the superconducting loop size). Phase jumps as shown in Figs. \ref{fig:spectroscopyinB}(d) and \ref{fig:spectroscopyinB}(e) are actually visible in virtually every spectroscopic measurement of the ABS structure in planar junctions \cite{PhysRevLett.124.226801, Moehle2022, PhysRevLett.130.116203, Fornieri2019, PhysRevB.107.245304,PhysRevLett.130.096202}. For instance, in the results of Ref. \cite{Moehle2022}, upon closer examination of the energy phase relation, it was found that the measured Andreev bound states exhibit prominent cusps, a phenomenon not anticipated for transparent junctions. These cusps signify the occurrence of phase slips, which manifest themselves when the superconducting loop that embeds the junction possesses a substantial inductance. The phase slips in the experiment of Ref. \cite{Moehle2022} showed that it is possible to probe only a part of the ABS spectrum, which then affected the outcome of the measurement in a stronger {\it perpendicular field}, where the asymmetry in local spectroscopy measured from top and bottom was obtained. As these types of structures \cite{PhysRevLett.124.226801, Moehle2022} are in principle to be used for nonlocal spectroscopy, the observed limited phase space probing represents a serious obstacle in probing the topological transition in them.

Let us discuss the conditions that will make it more favorable to bias the junction with phase $\pi$. Since it is the $LI(\phi)$ term that induces the non-linearity of flux to phase conversion, one should consider limiting it to restore the possibility of realizing the $\pi$ phase difference at small Zeeman fields. We approximate the condition under which increasing $B_{\perp}$ results in linear growth of the phase in the vicinity of the values of $\pi$ (mod $2\pi$) as when the second local maximum of $B_{\perp}(\phi)$ becomes larger than the first in each repeating $2\pi$ segment. For the case of $\tau \rightarrow 1$ we can analytically estimate the values of these two extrema, which leads to the condition
\begin{equation}
\varphi_\sigma + \frac{2\pi}{\Phi_0}LI_+(\pi + 2\varphi_\sigma) < 0,
\end{equation}
where $I_+$ is half of the total current, assuming the same number of spin positive and spin negative modes. Solving it for $L$ and $B_{\perp}$ yields the critical magnetic field at which $\pi$ phase biasing becomes possible. We plot the resulting diagram in the inset of Fig. \ref{fig:analyticalBtophase}(b) where blue denotes the parameter range that allows one to obtain the phase bias $\pi$. We observe a rapid growth of the critical field with an increase in $L$. The inductance of the superconducting loop is typically dominated by the kinetic inductance $L = l\hbar R_0/w\pi\Delta$ \cite{Annunziata_2010} where $l$ is the length, $w$ is the width of the arm of the superconducting loop and $R_0$ is normal state sheet resistance \cite{PhysRevB.107.245304}. Therefore, smaller loops with wide arms could in principle be used to decrease the critical Zeeman field, which is necessary to phase bias the junction into the topological regime.

Limiting the current is typically less favorable because it requires either limiting the transparency of the junction by decreasing the mean free path or making the width of the junction ($W_j$) smaller, thus decreasing the number of ABS. The latter is again unfavorable because it induces overlap between Majorana modes, lifting their degeneracy and would require usages of extended geometries \cite{PhysRevLett.125.086802, PhysRevB.104.155428}. An alternative approach could involve applying the gate voltage to the normal region of the junction, which in principle could decrease the number of ABS and therefore decrease the current in the junction \cite{Prasanna_note}.

\section{Summary and conclusions}
In this theoretical study, we investigated the possibility of detection of the topological transition in a planar Josephson junction via the nonlocal spectroscopy technique. We showed that the topological transition that is associated with the fermion parity change which is controlled by the in-plane magnetic field and the phase difference in the junction results in a change of the sign of the nonlocal conductance at zero energy. We showed that this phenomenon is directly related to the change in the quasiparticle character of the bands and can be used to determine the topological transition in the transport measurements. As we showed, in a realistic situation the control of the phase bias in the junction is strongly dependent on the strength of the in-plane magnetic field as the Zeeman interaction controls the current-phase relation. This leads to the inability of scanning the entire phase space, specifically reaching the $\pi$ bias required for topological transition at a small Zeeman interaction strength, unless the inductance of the superconducting loop embedding the junction or the current in the junction is considerably reduced.

\section{Acknowledgement}
We acknowledge the stimulating discussions with S. Goswami, C. M. Moehle, P. K. Rout and N. A. Jainandunsing. This work was supported by the National Science Center, Poland (NCN), Agreement No. UMO-2020/38/E/ST3/00418.

\bibliography{Ref}
\end{document}